\documentstyle[epsfig,twocolumn,prl,aps]{revtex}
\begin{document}
\draft
\twocolumn[\hsize\textwidth\columnwidth\hsize\csname @twocolumnfalse\endcsname
\title{Dynamics of 
Nonlocality for A Two-Mode Squeezed State in Thermal Environment} 
\author{Hyunseok Jeong $^1$\footnote{jeonghs@quanta.sogang.ac.kr}, Jinhyoung
  Lee$^1$\footnote{hyoung@quanta.sogang.ac.kr}, and
  M. S. Kim$^{1,2}$
} 
\address{$^1$ Department of Physics, Sogang University, 
CPO Box 1142, Seoul 100-611, Korea \\
$^2$Department of Applied
    Mathematics and Theoretical Physics, The Queen's University of 
    Belfast, Belfast BT7 1NN, Northern Ireland}
\date{\today} 

\maketitle

\begin{abstract}
  We investigate the time evolution of nonlocality for a two-mode
  squeezed state in the thermal environment. The initial two-mode pure
  squeezed state is nonlocal with a stronger nonlocality for a larger
  degree of squeezing. It is found that the larger the degree of
  initial squeezing is, the more rapidly the squeezed state loses its
  nonlocality.  We explain this by the rapid destruction of quantum
  coherence for the strongly squeezed state.
\end{abstract}
\vspace{2truecm}

\vskip1pc] 

\newpage

\section{INTRODUCTION}
\label{sec1}

Quantum nonlocality is one of the most profound features of quantum
mechanics\cite{EPR,Bell}. It enables current developments of quantum
information theory encompassing quantum
teleportation\cite{Bennett93,Ndim,Vaidman,Braunstein,Ralph,Furusawa},
quantum computation\cite{Deutsch85}, and quantum cryptography
\cite{Ekert91}.  There have been studies on tests of quantum
nonlocality versus local realism.  Bell suggested an inequality which
any local hidden variable theory must obey \cite{Bell}. Several types
of Bell's
inequalities were derived in terms of two-body correlation functions of
two measurement variables at distant places \cite{Clauser}
for the test of quantum nonlocality of a spin-1/2 or
SU(2) system.

A spin-1/2 system can be utilized as a qubit for quantum computation.
Quantum nonlocality of the spin-1/2 system is required as a quantum
channel to teleport an unknown qubit state \cite{Bennett93,Boschi98}.
In fact, it is possible to teleport not only a two-dimensional
spin-1/2 quantum state but also a $N$-dimensional state \cite{Ndim}
and a continuous-variable state \cite{Vaidman}.  The type of the
quantum channel depends on the dimension of Hilbert space of an
unknown quantum state.  For the teleportation of a continuous-variable
state, the quantum channel should be in an entangled
continuous-variable state such as the two-mode squeezed state
\cite{Braunstein}. Recently, practical implementation of the quantum
teleportation for the continuous-variable state has been realized
experimentally using the two-mode squeezed field
\cite{Ralph,Furusawa}.  In the quantum teleportation, the most
important ingredient is the quantum nonlocality of the channel which
can be easily destroyed in the nature.  In this paper
we are interested in how the
thermal environment affects the quantum nonlocality of the two-mode
squeezed field.

Quantum nonlocality of an entangled continuous-variable state has been
discussed using the Schmidt form for entangled nonorthogonal states
\cite{Mann95} and the quadrature-phase homodyne measurement
\cite{Munro}.  A given state is nonlocal when it violates any Bell's
inequality.  In fact, a state does not have to violate all the
possible Bell's inequalities to be considered quantum nonlocal.  A
state is quantum nonlocal for the given Bell's inequality which is
violated by the measurement of the state.  Banaszek and W\'{o}dkiewicz
defined Bell's inequality based on the parity measurement and they
found that the two-mode squeezed state violates the Bell's inequality,
showing quantum nonlocality \cite{Banaszek}.

In this paper we study the dynamic behavior of the 
quantum nonlocality based on the parity measurement for
the two-mode squeezed state in the thermal environment.  A measurement
of the degree of quantum nonlocality is defined here by the maximal
violation of Bell's inequality.  The nonlocality is stronger for the
squeezed state with a larger degree of squeezing.  It is found that
the nonlocality disappears more rapidly in the thermal environment 
as the initial state is squeezed more.

This paper is organized as follows. In Sec.~\ref{sec2}, Bell's
inequality based on the parity measurement is discussed.  The parity
measurement is directly related to the Wigner function.  In
Sec.~\ref{sec3}, the two-mode master equation is solved for the
dynamics of the Wigner function of the initial two-mode squeezed
state.  The convolution theory is utilized in the solution \cite{MSK}.
We investigate the dynamic behavior of the quantum nonlocality
measured by the maximum violation of Bell's inequality for the
two-mode squeezed state in the thermal environment in Sec.~\ref{sec4}.

\section{Bell's inequality by parity measurement}
\label{sec2}

It is important to choose the type of measurement variables when
testing nonlocality for a given state. In the original EPR gedanken
experiment \cite{EPR}, Einstein, Podolsky, and Rosen considered the
positions  (or the momenta) of two 
particles as the measurement variables to discuss  
the two-body correlation.  Bell \cite{Bell} argued 
that the EPR wave function does
not exhibit nonlocality because its Wigner function
$W(x_1,p_1;x_2,p_2)$ is positive everywhere, allowing the description
by a local hidden variable theory. Munro showed that various types of
Bell's inequalities are not violated in terms of
the homodyne measurements of two particles \cite{Munro,Ou92}.
To the contrary, Banaszek and W\'{o}dkiewicz \cite{Banaszek} examined
even and odd parities 
as the measurement variables and showed
that the EPR state and the two-mode squeezed state are nonlocal in the
sense that they violate Bell's inequalities such as Clauser and Horne
inequality and Clauser-Horne-Shimony-Holt inequality.

The even and odd parity operators, $\hat{O}_e$ and $\hat{O}_o$, are the
projection operators to measure the probabilities of the field having
even and odd numbers of photons, respectively:
\begin{equation}
\label{even-operator}
\hat{O}_e = \sum_{n=0}^\infty |2n\rangle \langle 2n|~~~;~~~
\hat{O}_o = \sum_{n=0}^\infty |2n+1\rangle \langle 2n+1|.
\end{equation}
The Wigner function at the origin of phase space for a state of the
density operator $\hat{\rho}$ is proportional to the mean parity
\cite{Moya93}:
\begin{equation}
\label{Wigner-origin}
W(0)=(2/\pi)\mbox{Tr}\left[(\hat{O}_e-\hat{O}_o)\hat{\rho}\right].
\end{equation}
Further, the Wigner function $W(\alpha)$ at the phase point $\alpha$
is the mean parity for the displaced original state
\begin{equation}
\label{Wigner-alpha}
W(\alpha)=(2/\pi)\mbox{Tr}[(\hat{O}_e-\hat{O}_o)
\hat{D}(\alpha)\hat{\rho}\hat{D}^\dag(\alpha)]
\end{equation}
where $\hat{D}(\alpha)$ is the displacement operator \cite{Moya93}.

So far the argument has been confined to the parity measurement 
of a single-mode field.  As the quantum nonlocality can be  
discussed for two-mode fields, we thus define 
the quantum correlation operator based on the joint parity
measurements:
\begin{eqnarray}
  \label{eq:tpcf}
  \hat{\Pi}^{ab}(\alpha, \beta) =
  \hat{\Pi}_e^{a}(\alpha)\hat{\Pi}_e^{b}(\beta)
 -\hat{\Pi}_e^{a}(\alpha)\hat{\Pi}_o^{b}(\beta) \nonumber \\
 -\hat{\Pi}_o^{a}(\alpha)\hat{\Pi}_e^{b}(\beta)
 +\hat{\Pi}_o^{a}(\alpha)\hat{\Pi}_o^{b}(\beta)
\end{eqnarray}
where the superscripts $a$ and $b$ denote the modes and
the displaced parity operator, $\Pi_{e,o}(\alpha)$, 
is defined as
\begin{equation}
  \label{eq:dpo}
  \hat{\Pi}_{e,o}(\alpha) = \hat{D}(\alpha) \hat{O}_{e,o} 
  \hat{D}^\dagger(\alpha).
\end{equation}
The displaced parity operator acts like a rotated spin
projection operator in the spin measurement.
We can easily 
derive that the local hidden variable theory imposes
the following Bell's inequality \cite{Banaszek}
\begin{eqnarray}
\label{inequality1}
|B(\alpha,\beta)|\equiv
|\langle \hat{\Pi}^{ab}(\alpha, \beta)+\hat{\Pi}^{ab}(\alpha, \beta')\nonumber\\
+\hat{\Pi}^{ab}(\alpha', \beta)-\hat{\Pi}^{ab}(\alpha', \beta')
\rangle|\leq 2
\end{eqnarray}
where we call $B(\alpha,\beta)$ as the Bell function. 

By a simple extension of the relation (\ref{Wigner-alpha}), the
two-mode Wigner function is found to be proportional to the mean of
$\hat{\Pi}_{ab}$ such that $W(\alpha,\beta) = (4/\pi^2)$
Tr$[\hat{\rho}_{ab}\hat{\Pi}^{ab}(\alpha, \beta)]$ for the two-mode
state of the density operator $\hat{\rho}_{ab}$.  The Bell
function (\ref{inequality1}) can then be written in terms of the
Wigner functions at different phase-space points,
\begin{equation}
  \label{eq:DefofB}
B(\alpha, \beta) = \frac{\pi^2}{4} \left[W(0,0) + W(\alpha,0) +
  W(0,\beta) - W(\alpha,\beta)\right].  
\end{equation}
The type of Bell's inequality in Eq.~(\ref{inequality1}) 
was first discussed by Clauser, Horne, Shimony, and Holt
\cite{Clauser}.
Clauser and Horne later found another type of inequality which can
be also expressed in phase space using the quasiprobability $Q$
function \cite{Banaszek}.  The $Q$ function is related to 
the probability of the state having no photons.  The lower and
upper critical values of the Clauser-Horne Bell's inequality is -1 and 0.  
 
We have seen that the two-mode Wigner function is useful to test
quantum nonlocality of the given field so that, in the next section,
we find the evolution of the Wigner function for the initial two-mode
squeezed state coupled with the thermal environment.

\section{Time evolution of two-mode squeezed states in thermal
  environment}
\label{sec3}

The two-mode squeezed state is the correlated state of two field modes
$a$ and $b$ that can be generated by a nonlinear $\chi^{(2)}$ medium
\cite{Caves,Barnett}.  The two-mode pure squeezed state is obtained by
applying the unitary operator on the two-mode vacuum
\begin{equation}
  \label{eq:Densitymatrix}
|\Psi_{ab}(\sigma)\rangle =\exp\left(-\sigma \hat{a} \hat{b} + 
\sigma^* \hat{b}^\dagger
  \hat{a}^\dagger\right)|0_{a},0_{b}\rangle
\end{equation}
where $\sigma = s \exp(-i \varphi)$ and $\hat{a}$ ($\hat{b}$) is an
annihilation operator for the mode $a$ ($b$).  The value of $s$
determines the degree of squeezing.  The larger $s$ is, the more the
state is squeezed.

The Wigner function corresponding to the squeezed state is the Fourier
transform of its characteristic function $C_W(\zeta,\eta)$
\cite{Barnett},
\begin{equation}
  \label{eq:charfunc}
  C_W(\zeta,\eta) = {\rm Tr}\left\{\hat{\rho} \exp(\zeta
  \hat{a}^\dagger - \zeta^* \hat{a}) 
  \exp(\eta \hat{b}^\dagger - \eta^* \hat{b})\right\}.
\end{equation}
For the two-mode squeezed state of the density matrix $\hat{\rho} =
|\Psi_{ab}(\sigma)\rangle \langle \Psi_{ab}(\sigma)|$, 
the Wigner function is written
as
\begin{eqnarray}
  \label{eq:WignerofTMS}
W_{ab}(\alpha,\beta)= \frac{4}{\pi^2} \exp \Big[-2 \cosh(2s)
  \left(|\alpha|^2 + |\beta|^2\right) \nonumber\\+  2\sinh(2s) \left(\alpha \beta
  + \alpha^* \beta^*\right)\Big]. 
\end{eqnarray}
The correlated nature of the two-mode squeezed state is exhibited by
the $\alpha \beta$ cross-term which vanishes when $s=0$.

The Fokker-Planck equation (in Born-Markov approximation) describing
the time evolution of the Wigner function in the interaction picture
can be written as
\begin{eqnarray}
  \label{eq:MasterEq}
  \frac{\partial W_{ab}(\alpha,\beta,\tau)}{\partial\tau} = 
  {\gamma \over 2} \sum_{\alpha_i=\alpha,\beta} \Bigg
  [ \frac{\partial}{\partial 
      \alpha_i} \alpha_i + \frac{\partial}{\partial
      \alpha_i^*} \alpha_i^* \nonumber\\ + 2 \left( \frac{1}{2}+\bar{n} \right)
    \frac{\partial^2}{\partial \alpha_i \partial \alpha_i^*}\Bigg]
  W_{ab}(\alpha,\beta,\tau),
\end{eqnarray}
where we have assumed that the two modes of the environment are
independent each other and the energy decay rates of the two modes are
same and denoted by $\gamma$.  The two modes have the same average
thermal photon number $\bar{n}$. By solving the Fokker-Planck equation
(\ref{eq:MasterEq}), we get the time evolution of the Wigner function
at time $\tau$ to be given by the convolution of the original function
and the thermal environment \cite{MSK},
\begin{eqnarray}
  \label{eq:SolofMastereq}
W_{ab}(\alpha, \beta, \tau)=\frac{1}{t(\tau)^4} \int d^2 \zeta d^2 \eta
W_{a}^{th}(\zeta) W_{b}^{th}(\eta) \\ \times W_{ab}\left(\frac{\alpha-r(\tau)\zeta
 }{t(\tau)},\frac{\beta-r(\tau)\eta}{t(\tau)},\tau=0\right),
\end{eqnarray}
where the parameters $r(\tau)=\sqrt{1-e^{-\gamma\tau}}$ and
$t(\tau)=\sqrt{e^{-\gamma\tau}}$.  $W^{th}(\zeta)$ is the Wigner
function for the thermal state of the average thermal photon number
$\bar{n}$:
\begin{equation}
  \label{eq:Wignerofe}
W^{th}(\zeta)=\frac{2}{\pi (1+2 \bar{n})} \exp
  \left(-\frac{2|\zeta|^2}{1+2 \bar{n}}\right).
\end{equation}
Performing the integration in Eq.~(\ref{eq:SolofMastereq}), the
Wigner function for the initial two-mode squeezed state evolving in
the thermal environment is obtained as
\begin{eqnarray}
  \label{eq:Wignerweget}
W_{ab}(\alpha,\beta, \tau)={\cal N} \exp \Big[-E(\tau) (|\alpha|^2 +
|\beta|^2) \nonumber \\+ 
F(\tau) (\alpha \beta + \alpha^* \beta^*)\Big] 
\end{eqnarray}
where
\begin{eqnarray}
  \label{eq:Wignerweget2}
  E(\tau)&=&\frac{2r(\tau)^2(1+2\bar{n})+2t(\tau)^2\cosh2s}{D(\tau)}
  \nonumber \\
  F(\tau)&=&\frac{2 t(\tau)^2 \sinh 2s}{D(\tau)}\nonumber \\
  D(\tau)&=&t(\tau)^4 + 2 r(\tau)^2
  t(\tau)^2(1+2\bar{n})\cosh2s \nonumber \\ &&+r(\tau)^4(1+2\bar{n})^2
\end{eqnarray}
and ${\cal N}$ is the normalization factor.  In the limit of $s=0$,
the $\alpha \beta$-cross term vanishes and the state can be
represented by the direct product of each mode states such that
$W_{ab}(\alpha,\beta,\tau)= W_{a}(\alpha, \tau) W_{b}(\beta, \tau)$.
It is obvious that the Wigner function (\ref{eq:Wignerweget}) exhibits
the local characteristics in this limit.

The system will eventually assimilate with the environment which can
be seen in the Wigner function, at the limit of $\tau \rightarrow
\infty$,
\begin{equation}
  \label{eq:Wignerfinal}
W_{ab}(\alpha,\beta)=
\frac{4}{\pi^2(1+\bar{n})^2} \exp [-\frac{2}{(1+2
  \bar{n})}(|\alpha|^2+|\beta|^2)]. 
\end{equation}
This is the direct product of two thermal states in modes $a$ and
$b$.

\section{Evolution of quantum nonlocality}
\label{sec4}

Substituting Eq.~(\ref{eq:Wignerweget}) into Eq.~(\ref{eq:DefofB}), we
find the evolution of the nonlocality for the initial two-mode
squeezed state in the thermal environment.  The Bell function $B$ at time
$\tau$ is written by
\begin{eqnarray}
  \label{eq:Localweget}
B(\alpha, \beta, \tau) = \frac{\pi^2 {\cal N}}{4} \exp
\Big\{1 + \exp\left[- E(\tau) |\alpha|^2\right]~~~~ \nonumber \\+ \exp\left[-E(\tau)
    |\beta|^2\right]
          - \exp\big[- E(\tau) (|\alpha|^2+|\beta|^2)\nonumber\\ + 2 F(\tau)
          |\alpha \beta| \cos\theta\big]\Big\},~~~~~~~~~~~~~~~~~~~~~~~~~~~~~~~~~
\end{eqnarray}
where $\theta_{\alpha}$ and $\theta_{\beta}$ are the phases of
$\alpha$ and $\beta$ and $\theta=\theta_{\alpha}+\theta_{\beta}$.
When $\cos\theta=-1$, the Bell function $B_m(|\alpha|,|\beta|,\tau)$
is described by the absolute values $|\alpha|$ and $|\beta|$.  $B_m$
is symmetric in exchanging $\alpha$ and $\beta$ such that
$B_m(|\alpha|,|\beta|,\tau) = B_m(|\beta|,|\alpha|,\tau)$.  It is
straightforward to show that $B \le B_m$ at any instance of time
$\tau$. In order to find the evolution of the nonlocality, the maximal
value $|B|_{max}$ of the Bell function $B$ is calculated by the
steepest decent method \cite{numericalrecides} and using the
properties of $B_m(|\alpha|,|\beta|,\tau)$. {\it We say the field is
  quantum-mechanically nonlocal as $|B|_{max}$ is larger than 2 and
  the nonlocality is stronger as $|B|_{max}$ gets larger}.

The initial two-mode squeezed state is always nonlocal as
$|B|_{max}>2$ for $s > 0$. $|B|_{max}$ increases monotonously as the
degree $s$ of squeezing increases.  The state becomes maximally
nonlocal with $|B|_{max}\sim 2.19055$ as
$s\rightarrow\infty$\cite{Banaszek}.  In an intermediate time $0 <
\tau < \infty$, the pure squeezed state evolves to a two-mode mixed
squeezed state and nonlocality is lost at a certain evolution time.
Figs.~1 and 2 show $|B|_{max}$ versus the dimensionless time $r(\tau)$
defined in Eq.~(\ref{eq:SolofMastereq}).  
We find that the nonlocality initially
prepared persists until the characteristic time $\tau_c(s,\bar{n})$
depending on the temperature of the thermal environment and the
initial squeezing.  In Fig.~1 it is found that, when the environment
is the vacuum, $|B|_{max}$ decreases as time proceeds.  After reaching
at the minimum value, $|B|_{max}$ increases to 2 which is the value of
$|B|_{max}$ for the vacuum.  Even though it is not clearly seen in the
figure due to the scale of the figure, for any $\bar{n}\neq 0$ thermal
environment, $|B|_{max}$ increases to its value for the thermal field
after it decreases to a minimum.  In Fig.~1, as $\bar{n}$ gets larger
$|B|_{max}$ decreases much faster and further.

\begin{figure}[htbp]
  \begin{center}
    \includegraphics*[height=5.5cm,width=8cm]{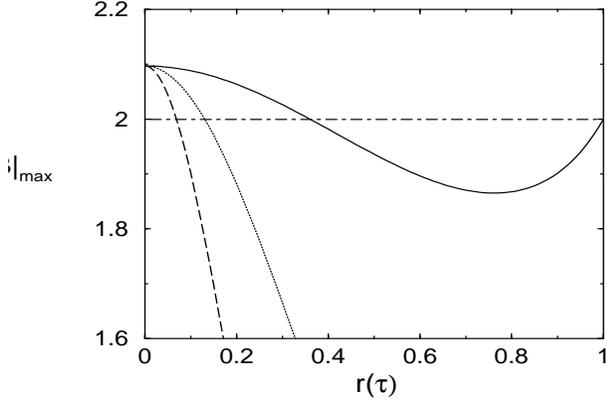}

    \caption{\small The time evolution of the maximal value $|B|_{max}$ of
      the Bell function versus the dimensionless time
      $r(\tau)\equiv\sqrt{1-\exp(-\gamma\tau)}$ which is 0 at $\tau=0$ and
      1 at $\tau=\infty$. The initial degree of squeezing $s=0.3$ and
      the average photon number $\bar{n}$ of the thermal environment
      is $\bar{n}=0$ (solid line), $\bar{n}=0.5$ (dotted line), and
      $\bar{n}=2$ (dashed line). The larger $\bar{n}$ is, the more
      rapidly the nonlocality is lost.}
  \end{center}
  \label{fig:2}
\end{figure}

In Figs.~2, we identify an interesting phenomenon that the larger the
initial degree of squeezing the more rapidly $|B|_{max}$ decreases.
We analyze the reason why
$|B|_{max}$ decreases more rapidly as the initial squeezing is larger
as follows.

The two-mode squeezed state (\ref{eq:Densitymatrix}) can be
represented by the continuous superposition of two-mode coherent
states  (A similar analysis was done for a
single-mode squeezed state \cite{Janszky})
\begin{equation}
\label{coh-sup}
|\Psi_{ab}(\sigma)\rangle = \int d^2\alpha G(\alpha,\sigma)
|\alpha,\alpha^*\mbox{e}^{i\varphi}\rangle
\end{equation}
where the Gaussian weight function 
\begin{equation}
\label{weght-function}
G(\alpha,\sigma) = (\pi \sinh s)^{-1}\exp\left[-\left({1-\tanh s
\over \tanh s}\right)|\alpha|^2\right].
\end{equation}
As $s$ gets larger, the weight of a large $\alpha$ state is greater so
that the contribution of $|\alpha,\alpha^*\mbox{e}^{i\varphi} \rangle$
of a large $\alpha$ becomes more important in the continuous
superposition (\ref{coh-sup}).

The quantum interference between coherent component states is the key
of quantum nature of the field.  The quantum interference is destroyed
by the environment.  The speed of destruction depends on the distance
between the coherent component states and the average thermal energy
of the environment \cite{Kim92}.  This is a reason why the macroscopic
quantum superposition state is not easily seen in the nature.  In the
continuous superposition (\ref{coh-sup}) we find that as the degree of
squeezing is larger, the superposition extends further so that the
quantum interference can be destroyed more easily.  
The quantum nonlocality in the
two-mode squeezed state is also originated from the quantum
interference between the coherent component states which can be
destroyed easily as the contribution of the large amplitude coherent
state becomes important.

\begin{figure}
  \begin{center}
    \includegraphics*[height=5.5cm,width=8cm]{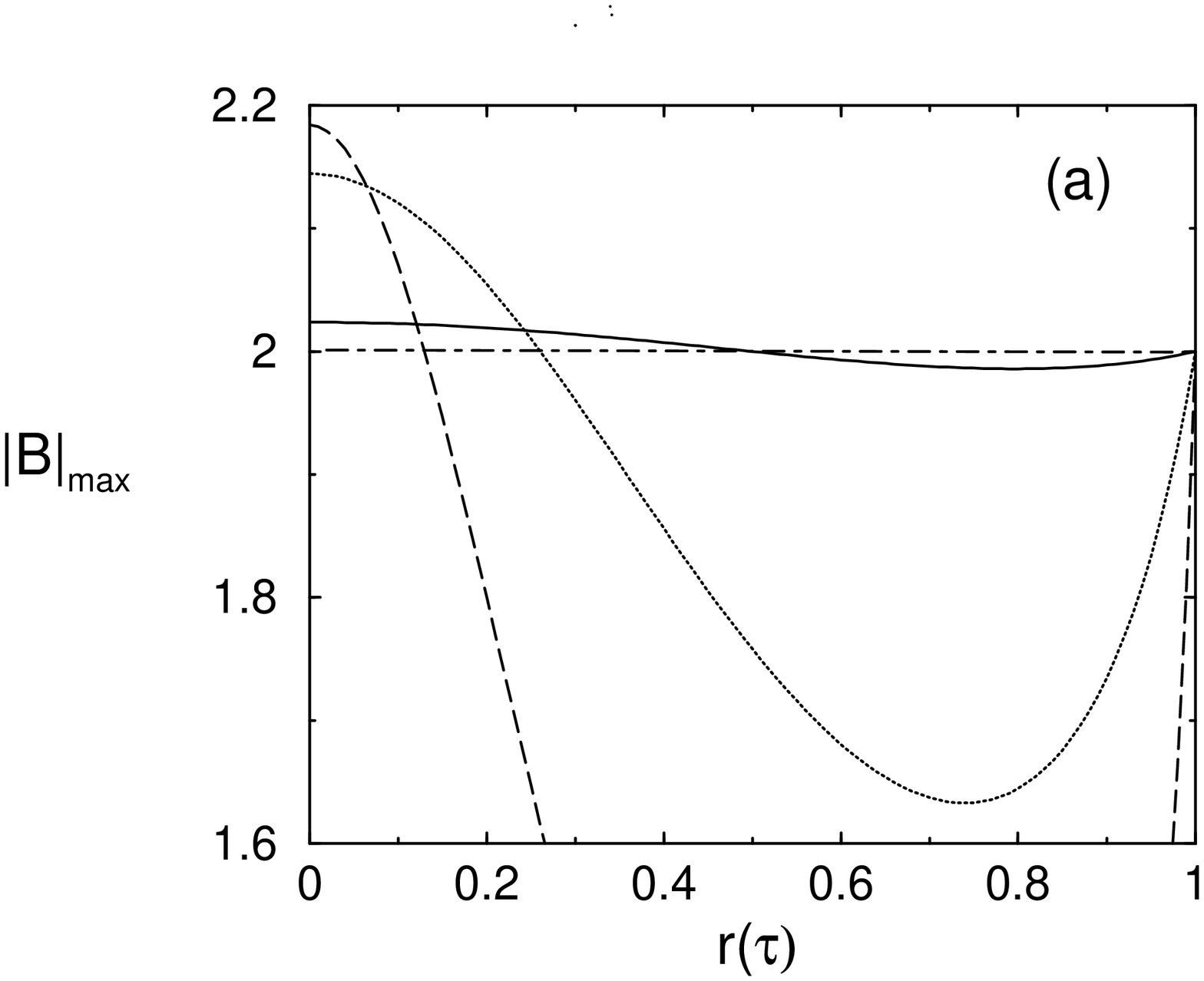}

    \includegraphics*[height=5.5cm,width=8cm]{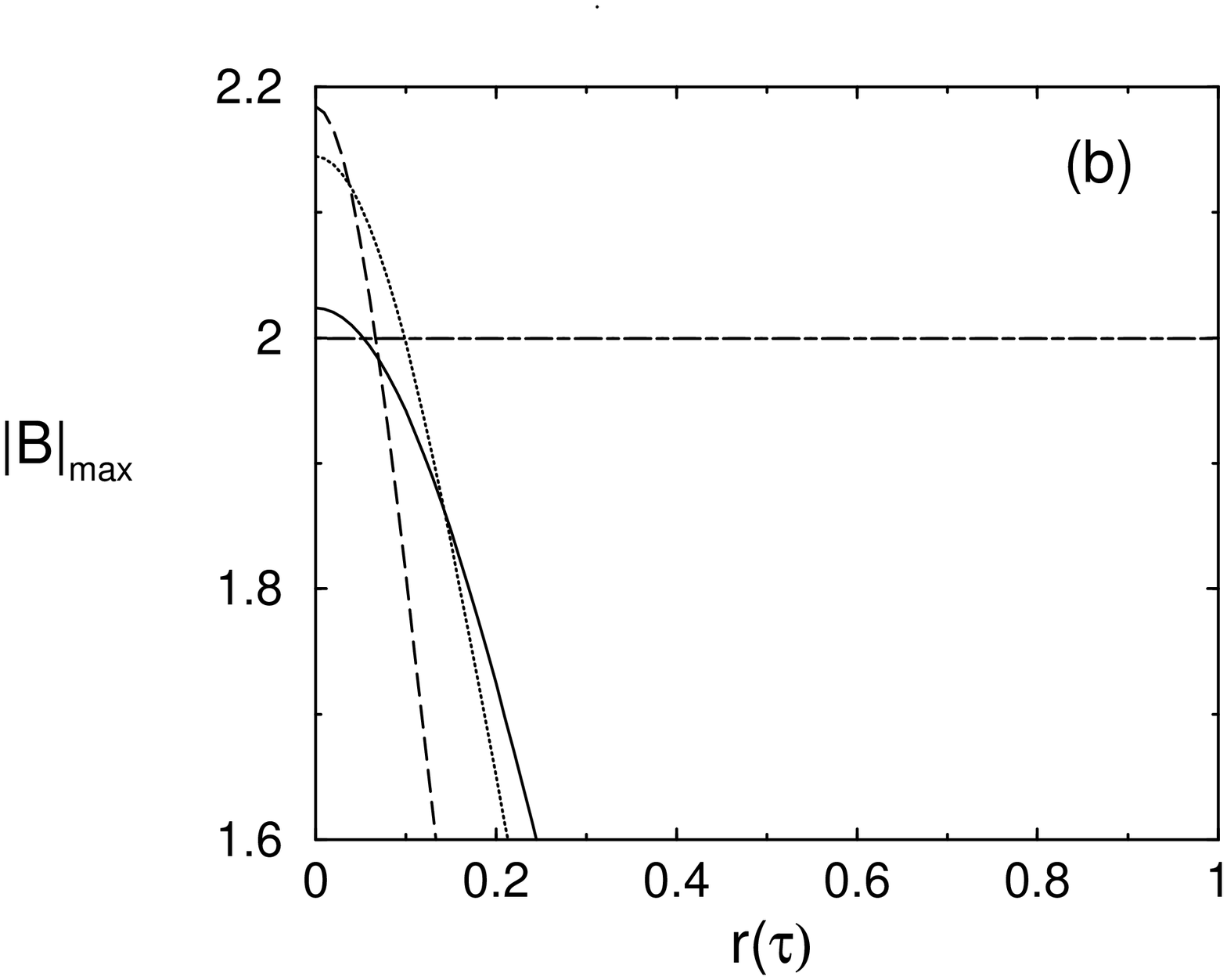}

    \caption{\small The time evolution of $|B|_{max}$ versus 
      $r(\tau) \equiv \sqrt{1-\exp(-\gamma\tau)}$ when the squeezed state is
      prepared with the initial degree of squeezing $s=0.1$(solid
      line), $s=0.5$(dotted line), and $s=1.0$(dashed line).  The
      two-mode squeezed state is coupled with the $\bar{n}=0$ vacuum
      (a) and the $\bar{n}=1$ thermal environment (b). In the vacuum,
      the larger the degree of squeezing is, the more rapidly the
      nonlocality is lost.  In the $\bar{n} =1$ thermal environment, we
      find that the nonlocality persists longer when the squeezing is
      $s \sim 0.5$.}
  \end{center}
  \label{fig:3}
\end{figure}

In fact the uncertainty increases to its maximum and decreases to the
value of the environment when a single-mode squeezed state is
influenced by the thermal environment \cite{Walls94}.  
The uncertainty increases
faster as the degree of squeezing is larger.  This can be explained
using the same argument as the lost of quantum nonlocality.
 
In Fig.~2(a), when the environment is in the vacuum, it is found that
the characteristic time $\tau_c(s,\bar{n})$ to lose the 
quantum nonlocality is shorter as the
initial degree of squeezing is larger.  
In Fig.~2(b), when the non-zero temperature 
thermal
environment ($\bar n \neq 0$) is concerned,  
we find that the larger degree of squeezing does not
necessarily result in the shorter characteristic time 
$\tau_c(s,\bar{n})$. 
This clearly shows that the characteristic time is a function of
the average number of thermal photons as well as the degree of squeezing.
However,
it is still true that $|B|_{max}$ decreases faster (the slope
of its curve is steeper) when $s$ is larger.  It is also found that
$|B|_{max}$ decreases faster for $\bar n \neq 0$ than for $\bar n =0$.

We have studied the dynamic behavior of the nonlocality for the
two-mode squeezed state in the thermal environment.  The two-mode
squeezed state can be used for the quantum channel in  quantum
teleportation of a continuous variable state.  The two-mode squeezed
state is found to be a nonlocal state regardless of its degree of
squeezing and the higher degree of squeezing brings about the larger
quantum nonlocality.  As the squeezed state is influenced by the
thermal environment the nonlocality is lost.  The rapidity of the loss
of nonlocality depends on the initial degree of squeezing and the
average thermal energy of the environment.  The more strongly the initial
field is squeezed, the more rapidly the maximum nonlocality decreases.
This has been analyzed extensively.

\acknowledgements 
We thank Professor J.W. Noh for bringing
Ref.\cite{Braunstein,Furusawa} to our attention.  This work is
supported by the BK21 Grant of the Korea Ministry of Education and
by the Sogang University Research Grants in 1999.


\begin{references} 

\bibitem{EPR} A. Einstein, B. Podolsky, and N. Rosen, Phys. Rev.
  {\bf 47}, 777(1935).

\bibitem{Bell} J. S. Bell, Physics {\bf 1}, 195 (1964).
  
\bibitem{Bennett93} C. H. Bennett, G. Brassard, C. Cr\'{e}peau, R.
  Jozsa, A. Peres, and W. K. Wootters, Phys. Rev. Lett. {\bf 70}, 1895
  (1993).  

\bibitem{Ndim} S. Stenholm and P. J. Bardroff, Phys. Rev. A {\bf 58},
  4373 (1998).

\bibitem{Vaidman} L. Vaidman, Phys. Rev. A. {\bf 49}, 1473(1994).
  
\bibitem{Braunstein} S. L. Braunstein and H. J. Kimble, Phys. Rev.
  Lett. {\bf 80}, 869(1998).

\bibitem{Ralph} T. C. Ralph and P. K. Lam, \prl {\bf 81}, 5668 (1998).

\bibitem{Furusawa} A. Furusawa, J. L. S\o rensen, S. L. Braunstein,
  C. A. Fuchs, H. J. Kimble and E. S. Polzik, Science {\bf 282}, 706(1998).

\bibitem{Deutsch85} D. Deutsch, Proc. R. Soc. Lond. A {\bf 400},97
  (1985).

\bibitem{Ekert91} A. K. Ekert, \prl {\bf 67}, 661 (1991).

\bibitem{Clauser} J. F. Clauser, M. A. Horne, A. Shimony, and R. A.
  Holt, Phys. Rev. Lett. {\bf 23}, 880(1969); 
  J. F. Clauser and M. A. Horne, \prd {\bf 10},
  526 (1974).
  
\bibitem{Boschi98} D. Bouwmeester, J.-W. Pan, K. Mattle, M. Eibl, 
  H. Weinfurter, Nature {\bf 390}, 575 (1997);
  D. Boschi, S. Branca, F. De Martini, L. Harcy, and 
  S. Popescu, \prl {\bf 80}, 1121 (1998).

\bibitem{Mann95} A. Mann, B. C. Sanders, and W. J. Munro, \pra
  {\bf  51}, 989 (1995).

\bibitem{Munro} B. Yurke and D. Stoler, \prl {\bf 79}, 4941 (1997);
  A. Gilchrist, P. Deuar, and M. D. Reid, \prl {\bf 80}, 3169
  (1998); W. J. Munro, \pra {\bf 59}, 4197 (1999).

\bibitem{Banaszek} K. Banaszek and K. W\'{o}dkiewicz, \pra {\bf 58},
  4345 (1998); K. Banaszek and K W\'{o}dkiewicz, \prl {\bf 82},
  2009(1999).

\bibitem{MSK} M. S. Kim and N. Imoto, Phys. Rev. A. {\bf 52}, 2401(1995).

\bibitem{Ou92} Z. Y. Ou, S. F. Pereira, H. J. Kimble, and K. C. Peng,
  \prl {\bf 68}, 3663 (1992).

\bibitem{Moya93} H. Moya-Cessa and P. L. Knight, \pra {\bf 48}, 2479
  (1993); B.-G. Englert, N. Sterpi, and H. Walther, Opt. Commun. 
  {\bf 100}, 526 (1993).

\bibitem{Caves} C. Caves and B. L. Schumacher, \pra {\bf 31}, 3093 (1985).
  
\bibitem{Barnett} S. M. Barnett and P. L. Knight, \jmo {\bf 34}, 841 (1987).

\bibitem{numericalrecides} W. H. Pres, B. P. Flannery, S. A.
  Teukolsky, and W. T. Vetterling, {\it Numerical Recipes}, (Cambridge
  University, Cambridge, 1988).

\bibitem{Janszky} J. Janszky and A. V. Vinogradov, \prl {\bf 64},
  2771 (1990); V. Bu\v{z}ek and P. L. Knight, Opt. Commun. {\bf 81},
  331 (1991).

\bibitem{Kim92} M. S. Kim and V. Bu\v{z}ek, \pra {\bf 46}, 4239 (1992).

\bibitem{Walls94} D. F. Walls and G. J. Milburn, {\em Quantum
    Optics} (Springer, Berlin, 1994) p.105. 

\end{references}
\end{document}